\documentclass[twocolumn,prl,aps]{revtex4}
\usepackage{graphicx}
\begin{document}

\title{Bohm's ``quantum potential" can be considered falsified by experiment}

\author{Antoine Suarez}
\address{Center for Quantum Philosophy \\ Ackermannstrasse 25, 8044 Zurich, Switzerland\\
suarez@leman.ch, www.quantumphil.org}

\date{October 8, 2014}

\begin{abstract}

A Michelson-Morley-type experiment is described, which exploits two-photon interference between entangled photons instead of classical light interference. In this experimental context, the negative result (no shift in the detection rates) rules out David Bohm's postulate of an infinite-speed time-ordered ``quantum potential", and thereby upholds the timeless standard quantum collapse.

\ \\
\textbf{Keywords:} Quantum nonlocality, Bohmian mechanics,  infinite-speed time-ordered quantum potential, relativity, timeless standard quantum collapse.

\end{abstract}

\pacs{03.65.Ta, 03.65.Ud, 03.30.+p}

\maketitle

\ \\
\textbf{1. Introduction}.\textemdash A main postulate of standard quantum mechanics is that the decision of the outcome happens at the moment of detection (``wavefunction collapse''). This implies a \emph{nonlocal} coordination of the detectors, which cannot be explained by influences propagating with velocity $v\leq c$. So early as 1927, at the 5$^{th}$ Solvay conference, Einstein objected to this postulate by means of a \emph{single-particle} gedanken-experiment. The quantum collapse, he argued, involves ``an entirely peculiar mechanism of action at a distance, which [...] implies to my mind a contradiction with the postulate of relativity." \cite{bv}.

Astonishingly Einstein's gedanken-experiment in 1927 has been first realized in 2012 by the single-photon experiment presented in \cite{Guerreiro12}. This experiment demonstrates nonlocally coordinated detector's behavior, and also highlights something Einstein did not mention: Nonlocality is necessary to preserve such a fundamental principle as energy conservation.\cite{Guerreiro12}

To avoid the standard quantum nonlocality at detection in single-particle experiments Einstein invoked in 1927 Louis de Broglie's explanation by means of ``particle and pilot wave" \cite{bv}, which implies that the outcome is determined at the beam-splitter.

Nonetheless, de Broglie's local picture cannot be extended to entanglement experiments with two or more particles. Einstein himself is at the origin of this insight with the celebrated EPR paper in 1935: The local hidden variables of de Broglie (particle and pilot wave) do not account for the nonlocal EPR correlations quantum mechanics predicts \cite{jb}. To overcome this problem David Bohm completed the de Broglie's local model with a ``nonlocal quantum potential''. The theory was quite inspiring for John Bell since it highlights that local hidden variable models do not suffice to explain quantum mechanics. Bell showed that such models fulfill the well known locality criteria of Bell inequalities, and are refuted by the experimental violation of these in the conventional Bell experiments \cite{jb}.

%%%%%%%%%%%%%%%%%%%%%%%%%%%%%%
\begin{figure}[t]
\includegraphics[trim = 10mm 185mm 10mm -6mm, clip, width=0.90\columnwidth]{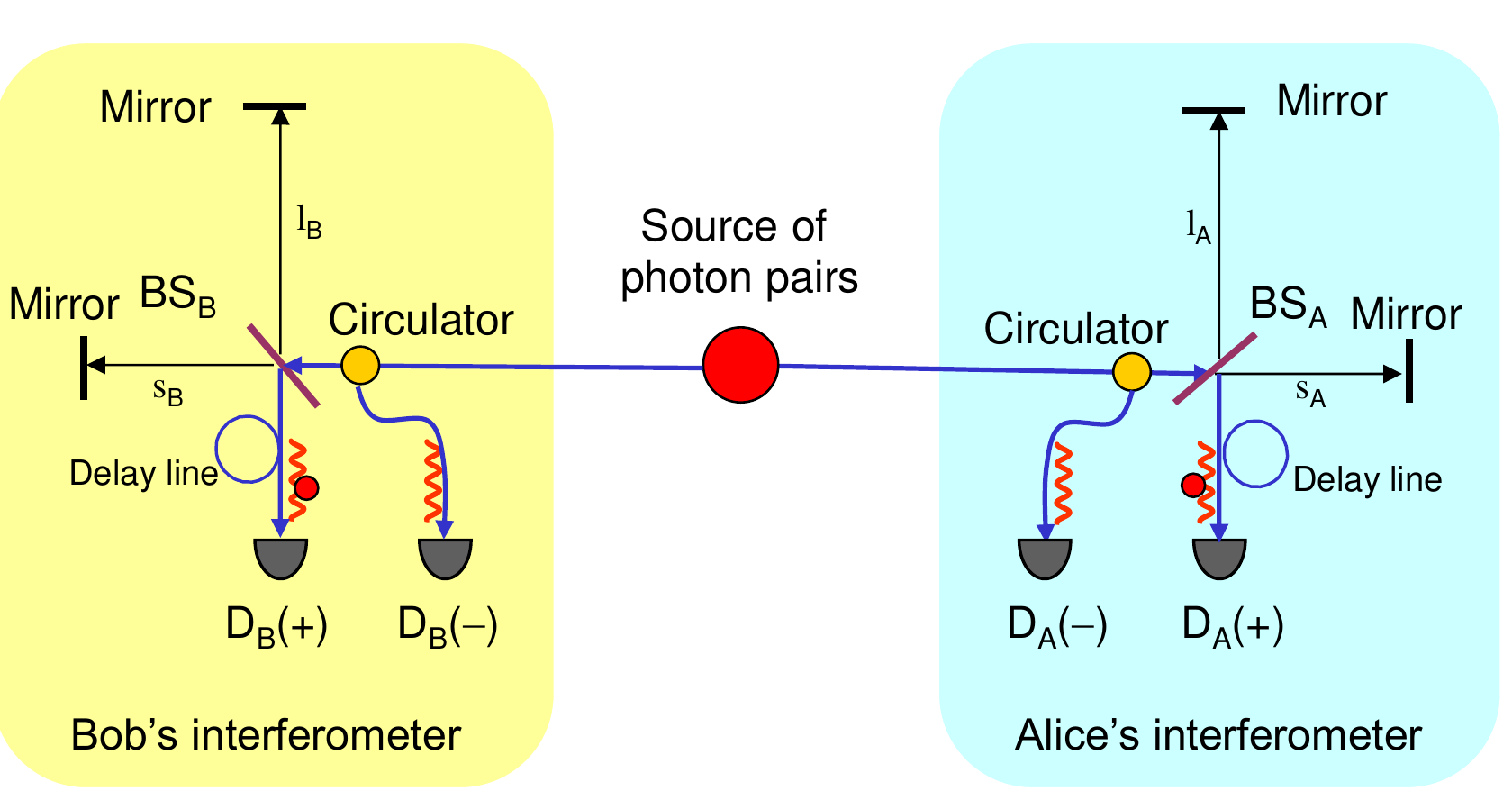}
\caption{\textbf{ Franson-type Bell experiment:} According to David Bohm the nonlocal correlations demonstrated by the experiment are produced by a ``quantum potential", that is, an influence between the outcome decisions at the beam-splitters BS$_{A}$ and BS$_{B}$, which works at infinite speed but nonetheless is time-ordered.}
\label{f1}
\end{figure}
%%%%%%%%%%%%%%%%%%%%%%%%%%%%%%

Moreover, the ``pilot wave'' is not a wave propagating in some material ether within the ordinary 3-space, but rather a mathematical entity defined in the so called ``3N-space or configuration space" (\cite{jb} p. 128). In other words, the ``wave'' guides the particle from outside the ordinary space-time. Therefore, if by ``realism" one intends the ontology assuming that all what matters for the physical reality is contained in the space-time, then de Broglie's description cannot be considered ``realistic" either.

This means that de Broglie-Bohm's model neither restores ``realism" nor ``locality" at the end. In particular, a main oddity of the model seems to be that the principle of nonlocality appears only in experiments with two or more particles but does not govern the single-particle experiments, whereas one would expect that so a fundamental feature like nonlocality pervades the whole quantum mechanics, and rules the single-particle phenomena too.

Probably the strongest argument in defence of the model is that it saves the idea of temporal causality: Bohm's ``quantum potential'', although traveling at infinite velocity, is supposed to produce time-ordered outcomes with relation to some ``preferred frame'' so that one outcome event is the cause, and the other the effect \cite{db,jb,aa}. The aim of this article is to show that this view leads to a prediction conflicting with relativity which can be tested, and argue that Bohm's time-ordered ``quantum potential" can actually be considered falsified by experiment on the basis of already available results.

\textbf{2. Time-ordered nonlocal quantum potential}.\textemdash Consider the Franson-type Bell experiment depicted in Figure \ref{f1}: A source produces pairs of entangled photons and each of them enters an unbalanced Michelson-Morley interferometer. Photon A (angular frequency $\omega_{A}$) enters Alice's interferometer to the right through the beam-splitter BS$_{A}$ and, after reflection in the mirrors, leaves through BS$_{A}$ again and gets detected. Photon B (angular frequency $\omega_{B}$) enters Bob's interferometer to the left through the beam-splitter BS$_{B}$ and, after reflection in the mirrors, leaves through BS$_{B}$ again and gets detected. The detectors are denoted D$_{A}(a)$ and D$_{B}(b)$ ($a, b\in\{+,-\}$). Each interferometer consists in a long arm of length $l_{i}$, and a short one of length $s_{i}$, $i\in\{A,B\}$. Frequency bandwidths and path alignments are chosen so that only the coincidence detections corresponding to the path pairs: $(s_{A},s_{B})$ and $(l_{A},l_{B})$ contribute constructively to the correlated outcomes in regions A and B, where $(s_{A},s_{B})$ denotes the pair of the two short arms, and $(l_{A},l_{B})$ the pair of the two long arms.

Suppose one of the measurements produces the value $a$ ($a\in\{+,-\}$), and the other the value $b$ ($b\in\{+,-\}$).
According to quantum mechanics the probability $P(a, b)$ of getting the joint outcome $(a, b)$ depends on the choice of the phase parameter $\Phi$ characterizing the paths or channels uniting the source and the detectors:

\begin{footnotesize}
\begin{eqnarray}\label{1}
    P(a=b)=\frac{1}{2}(1+\cos\Phi) \nonumber\\
    P(a\neq b)=\frac{1}{2}(1-\cos\Phi)
\end{eqnarray}
\end{footnotesize}

\noindent where $\mathit\Phi=\omega_{A}\tau_{A}+\omega_{B}\tau_{B}$ is the phase parameter, and $\tau_{i}$ the optical path difference i.e., the difference between the times light take to travel each path of the interferometer. If according to relativity one assumes the same light speed in the two paths of the interferometer, then $\tau_{i}=\frac{l_i - s_i}{c}$. The path lengths $l_i, s_i$ could be changed by means of a mobil mirrors.

Bell experiments, using two different values of $l_{A}$ and two different values of $l_{B}$, demonstrate that the correlations given by Equations (\ref{1}) violate locality criteria, the well known Bell's inequalities (see \cite{SZGS02} and references therein).

Suppose now that according to Bohm's ``nonlocal potential'' one describes the correlation between the outcome decisions at the two beam-splitters BS$_{A}$ and BS$_{B}$ as a time-ordered influence fulfilling the two following conditions:

a) it happens with infinite speed,

b) the decision at one of the beam-splitters happens first and determines (``is the cause of") the decision at the second beam-splitter.

The condition a) implies that the time-order invoked in b) can only come from the arrivals of the photons at the corresponding beam-splitters: So for instance in the experiment sketched in Figure \ref{f1}, one could suppose that after reflection in the mirrors Alice's photon  arrives at BS$_{A}$ before Bob's photon arrives at BS$_{B}$, and this way Alice's outcome is the cause and precedes Bob's one, which is the effect. But this assumption implies that \emph{it is possible to define the speed of light in an absolute way with relation to some ``preferred frame''}. And this prediction can be tested by using an experimental protocol similar to that used by Michelson-Morley 1887 as we show in the following section.

\begin{figure}
\includegraphics[trim = 2mm 80mm 0mm 6mm, clip, width=0.95\columnwidth]{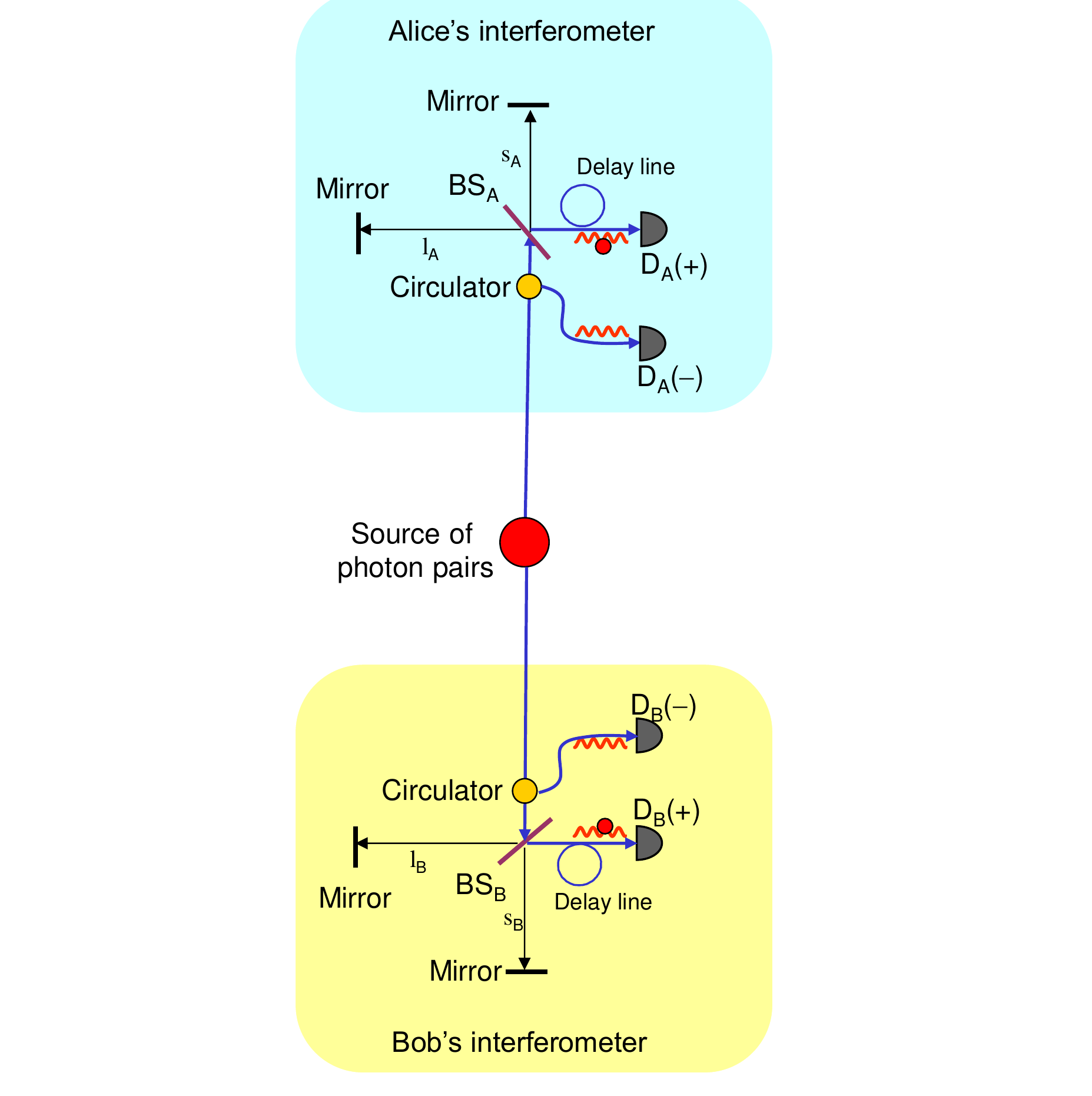}
\caption{\textbf{The Michelson-Morley entanglement experiment:} Setup rotated by $90\,^{\circ}$. The assumption of a time-ordered nonlocal quantum potential between the decision events at the beam-splitters implies that one can define the velocity of light in an absolute way with relation to a preferred referential frame, and this implies that the detection counting rates should change when the interferometer is rotated (see text).}
\label{f2}
\end{figure}

\ \\
\textbf{3. The  Michelson-Morley entanglement experiment}.\textemdash Consider the experiment of Figure \ref{f1} with the short arms oriented in the direction of the  Earth's orbital motion. We assume $l_i\approx l$ and $s_i\approx s$. According to the ``preferred frame'' assumption, if the Earth moves with velocity $v$ relative to the ``preferred frame'', the travel time of the light through each arm of Alice's interferometer exhibits a difference given by:

\begin{footnotesize}
\begin{eqnarray}
\tau_1= \frac{2sc}{c^2-v^2}-\frac{2l}{\sqrt{c^2-v^2}}\approx\frac{2s}{c}\left(1+\frac{v^2}{c^2}\right)-\frac{2l}{c}\left(1+\frac{v^2}{2c^2}\right)
\label{tau1}
\end{eqnarray}
\end{footnotesize}

By rotating the interferometer 90$^\circ$, as indicated in Figure \ref{f2}, one interchanges the orientation of the two paths relative to the ``preferred frame'' and gets the following travel time difference for the light:

\begin{footnotesize}
\begin{eqnarray}
\tau_2= \frac{2lc}{c^2-v^2}-\frac{2s}{\sqrt{c^2-v^2}}\approx\frac{2l}{c}\left(1+\frac{v^2}{c^2}\right)-\frac{2s}{c}\left(1+\frac{v^2}{2c^2}\right)
\label{tau2}
\end{eqnarray}
\end{footnotesize}

From (\ref{tau1}) and (\ref{tau2}) it follows that after rotation the total optical path difference of Alice's interferometer is given by:

\begin{footnotesize}
\begin{eqnarray}
\tau=\tau_1+\tau_2=(l+s)\frac{v^2}{c^3}
\label{tau}
\end{eqnarray}
\end{footnotesize}

Assuming $l_A\approx l_B$, and $s_A\approx s_B$ Equation (\ref{tau}) gives also the total optical path difference  of Bob's interferometer after rotation. Assuming also $\omega_{i}\approx\omega$ one gets the total phase shift:

\begin{footnotesize}
\begin{eqnarray}
\mathit\Delta\Phi&=&\omega_{A}\tau_{A}+\omega_{B}\tau_{B}\approx 2\omega(l+s)\frac{v^2}{c^3}\nonumber\\
&=&4\pi \nu (l+s) \frac{v^2}{c^3}=4\pi\frac{(l+s)}{\lambda}\frac{v^2}{c^2}
\label{Phi}
\end{eqnarray}
\end{footnotesize}

\noindent where $\nu$ means the frequency and $\lambda$ the wavelength of the photons.

Therefore, the assumption of the time-ordered nonlocal quantum potential leads to the prediction that the rotation of 90$^\circ$ of the setup (Figure \ref{f2}) produces a phase shift given by (\ref{Phi}), which should change the probabilities in Equation (\ref{1}) and the corresponding coincidence counting rates of the detectors.

Like in the Michelson-Morley experiment \cite{MM} we take $v=30 km/s$ as the orbital velocity of the Earth around the Sun \cite{mm}, and assume that at some moment of the day (because of the Earth rotation) one of the arms will be oriented near to the direction of the ``preferred frame''. We assume photons of wavelength about $\lambda=1550nm$.

We try to get a well testable prediction by setting parameters such that the counting rate changes from $P(a=b)=0.50$ to $P(a=b)=0.25$ respectively from $P(a\neq b)=0.50$ to $P(a\neq b)=0.75$ when the interferometer is rotated by 90$^\circ$. According to (\ref{1}) this counting rate shift corresponds to a phase shift of $\Delta\mathit\Phi=\frac{\pi}{6}$, and from (\ref{Phi}) one gets the following length $l+s$:

\begin{footnotesize}
\begin{eqnarray}
4\pi\frac{(l+s)}{\lambda}\frac{v^2}{c^2}=\frac{\pi}{6}\;\; \Rightarrow \;\;\; (l+s) = 6.25 m
\label{MM3}
\end{eqnarray}
\end{footnotesize}

Hence, if one uses a setup fulfilling (\ref{MM3}), the ``preferred frame'' theory predicts that at some moment of the day the counting rates of the detectors fulfill:

\begin{footnotesize}
\begin{eqnarray}
&&P(a=b)=0.50,\;\;\; \texttt{before rotation of 90$^\circ$} \nonumber\\
&&P(a=b)=0.25,\;\;\;\texttt{after rotation of 90$^\circ$}
\label{MM4}
\end{eqnarray}
\end{footnotesize}

Equation (\ref{MM4}) provides a clear trial of Bohm's ``time-ordered nonlocal quantum potential'' assumption.

\ \\
\textbf{4. Discussion}.\textemdash David Bohm's assumption of an ``infinite-speed time-ordered quantum potential'' is generally supposed to reproduce the experimental predictions of quantum mechanics, and, so far, considered a possible causal alternative to the standard interpretation of the timeless wavefunction collapse at detection (see for instance \cite{jb,aa,grö,cb,sbsg}). Strictly speaking, Bohm's time-ordered quantum potential implies disappearance of the quantum correlation in case the decisions at the beam-splitters BS$_{A}$ and BS$_{B}$ happen simultaneously in the assumed ``preferred frame", and hence it is actually at odds with standard quantum physics \cite{as09}. Nonetheless this prediction cannot be tested by a real experiment.

By contrast, in the experiment presented in the preceded section Bohm's assumption implies the shift in the counting rates predicted by (\ref{MM4}). This prediction conflicts with relativity and \emph{is} testable. Although a real experiment would be ``nice to have", it does not seem required if one considers that the falsification of the prediction (\ref{MM4}) results by induction from the Michelson-Morley experiments repeatedly performed in the past. To this extent the negative result of these experiments can be straightforwardly extended to the entanglement version presented in the precedent section to conclude that shift predicted by (\ref{MM4}) will not be observed. And this means that the Michelson-Morley entanglement experiment (Figures \ref{f1} and \ref{f2}) rules out Bohm's ``infinite-speed time-ordered quantum potential''.

Regarding the standard quantum collapse at detection does not invoke any time-order. The negative result in the experiment of the Figures \ref{f1} and \ref{f2} upholds this standard view of a timeless quantum collapse: Indeed, if the ``time-ordered quantum potential" is refuted, then the idea of a timeless (non time-ordered) infinite-speed coordination between the beam-splitters can be considered conceptually equivalent to the standard quantum collapse involving coordination between the detectors \cite{as14}. In turn, confirmation of relativity by the experiment shows that the quantum correlations cannot be explained by influences propagating with velocity $v\leq c$; the timeless quantum collapse implies relativity, and the relativistic structure of the space-time implies that the coordinated behavior of the two detectors D$_{A}(a)$ (or the two detectors D$_{B}(b)$) requires nonlocal coordination. Hence nonlocality at detection is a fundamental feature of quantum physics which appears already in single-particle experiments \cite{Guerreiro12}, while Bell's nonlocality makes sense only in experiments with two or more particles \cite{as14}.

Furthermore it has already been pointed out \cite{sbsg} that if one assumes nonlocality at detection, then any disappearance of nonlocal coordination between the detectors violates straightforwardly the conservation of energy in the single quantum events and, consequently, all other testable alternative models proposed to date are ruled out by the experiment in \cite{Guerreiro12}. Therefore the Michelson-Morley entanglement experiment presented in this article by ruling out Bohm's ``time-ordered quantum potential" and confirming relativity upholds thoroughly standard quantum mechanics.

\textbf{5. Conclusion}.\textemdash It is noteworthy that Bohmian mechanics conflicts with both, standard quantum mechanics and relativity. Whereas the conflict with quantum mechanics is not testable, the conflict with relativity can be tested through the experiment we have presented in this paper. Therefore Bohm's `` preferred frame" assumption can be considered falsified by experiment to the same extent as relativity is considered to be confirmed by it.

By contrast, the standard quantum collapse at detection ignores the ``preferred frame'' (time-order) and thereby implicitly contains relativity. The proposed experiment confirms this view and highlights that relativity and quantum physics are two inseparable aspects of one and the same description of the physical reality. These two theories neither are incompatible with each other nor have a ``frail peaceful coexistence'', but rather imply each other: we can't have one without the other.

\ \\
\emph{Acknowledgments}: I am thankful to  Andr\'{e} Stefanov and B\"{a}nz Bessire for insightful comments, and acknowledge discussions with Nicolas Brunner, Bruno Sanguinetti, and Hugo Zbinden.

\end{document}